\documentclass[floatfix,superscriptaddress]{revtex4}
\usepackage[T1]{fontenc}
\usepackage[latin9]{inputenc}
\usepackage{color}
\usepackage{soul}
\usepackage{multirow}
\usepackage{relsize}
\usepackage{graphicx}
\usepackage{calc}
\usepackage{amsmath}
\usepackage{babel}
\usepackage{float}
\usepackage{svg}
\usepackage{booktabs}
\usepackage{comment}
\usepackage{lipsum}
\usepackage{float}
\usepackage[dvipsnames]{xcolor}
\usepackage{graphicx}
\usepackage{dcolumn}
\usepackage[normalem]{ulem}
\makeatother
\usepackage{siunitx}
\usepackage[flushleft]{threeparttable}
\usepackage{xargs}                     

\graphicspath{{figs/}}
\makeatletter

\usepackage{tikz}

\usetikzlibrary{arrows.meta,positioning,calc,fit}
\usepackage{comment}
\usepackage{tabularx}
\usetikzlibrary{arrows.meta,positioning,calc,decorations.pathmorphing,decorations.markings}
\begin{document}
\title{Short-Pulse High-Power THz Generation Using Optical Klystron FELs: Simulation Results}
\author{Najmeh Mirian}\email{n.mirian@hzdr.de}

\affiliation{Helmholtz-Zentrum Dresden-Rossendorf HZDR, 01328 Dresden Germany}
\date{\today}

\begin{abstract}
The generation of high-power radiation in the terahertz (THz) regime using free-electron lasers (FELs) is challenging due to strong diffraction and pronounced slippage effects. These constraints often limit the achievable pulse duration and peak power in conventional single-pass THz FEL configurations. In this work, we investigate an unseeded optical klystron (OK) FEL concept tailored for the THz regime.
 Using time-dependent three-dimensional simulations for resonant wavelengths of 10, 30, and 100 $\mu$m,
we demonstrate that this approach enables the generation of coherent ultrashort THz pulses with durations of a sub picoseconds (FWHM) and peak powers in the multi-hundred-megawatt range at wavelengths 10 and 30 $\mu$m. 
To address the slippage challenge, we propose and numerically demonstrate a novel chicane-embedded optical delay scheme, which restores phase alignment between the radiation and  microbunched electrons. Simulations confirm that careful tuning of the dispersive strengths allows staged amplification, preserving beam quality and reaching multi-megawatt output power. These results highlight the potential of THz-tailored optical klystrons to generate compact, short, and high-intensity THz pulses, and lay the groundwork for future experimental studies and facility implementation.
\end{abstract}

\maketitle
\section{Introduction}

Free-electron lasers (FELs) provide widely tunable, high-brilliance radiation covering wavelengths from the terahertz (THz) range to hard X-rays~\cite{Kubarev2006,Neil2000,Emma2010}. 
Unlike conventional lasers, FELs exploit the interaction of a relativistic electron beam with a periodic magnetic field in an undulator, where the induced energy modulation evolves into a density modulation at the radiation wavelength. 
This microbunching enables coherent emission and, under suitable conditions, exponential amplification of the radiation field. 
While the self-amplified spontaneous emission (SASE) FEL~\cite{Emma2010} has been highly successful and is now the standard for short-wavelength FELs, it also has intrinsic limitations: start-up from shot noise degrades temporal coherence, the required undulator length can be very large, and accumulated slippage between the electron bunch and the radiation field affects the temporal structure of the output pulse~\cite{Madey1971,Bonifacio1984}.

At long wavelengths, the slippage length becomes comparable to or even larger than the electron bunch length, which makes the overlap between the optical field and the electron beam difficult to maintain. Consequently, relatively long electron bunches are required to achieve stable FEL amplification in THz range. For example, at the PITZ facility at DESY, bunches of about 15-20~ps duration were required to sustain THz FEL emission at $\lambda_r \sim 100~\mu$m~\cite{Krasilnikov}.

Several strategies have been explored to address these challenges.
One common approach is to operate THz FELs in waveguide-assisted configurations, where the transverse confinement of the radiation enhances the coupling between the electron beam and the electromagnetic mode, partially compensating for the reduced FEL parameter at long wavelengths and low beam energies. Such schemes have enabled compact THz FELs with reasonable gain lengths but are often limited by waveguide losses, mode competition, and wakefield effects at high charge \cite{Miginsky2015, Yang2024,Fisher2022}.

Alternative concepts exploit superradiant or coherent spontaneous emission regimes, where the electron bunch length is comparable to or shorter than the radiation wavelength, allowing the FEL to operate without exponential gain \cite{Fisher2024, Chulkov2014, Liang2026}. While attractive for generating high peak power in the THz range, these schemes typically require very short bunches and offer limited spectral tunability and scalability.
Despite these advances, existing approaches remain constrained by undulator length, beam quality preservation, or operational flexibility, motivating the exploration of alternative FEL concepts tailored to the specific challenges of the THz regime.

An alternative approach is the \textit{optical klystron} (OK), originally proposed as an optical analogue of the microwave klystron and first applied in the context of FEL oscillators \cite{Vinokurov1971,Orzechowski1986}. 
In its simplest form, the OK consists of two undulators separated by a dispersive chicane: the first undulator induces a weak energy modulation, which is converted into density modulation by the chicane, enabling coherent radiation emission in the second undulator. A comprehensive historical overview of the optical klystron concept, from its original proposal to its subsequent theoretical and experimental developments, is presented in Ref.\cite{Penco2015}.

While the optical klystron concept was initially developed for oscillator configurations, it has later been extended to enhance single-pass high-gain FELs operating in the SASE regime \cite{Ding2006,Penco2015}.
In this case, the pre-bunching provided by the optical klystron leads to an increased initial bunching factor and a reduced effective gain length, resulting in earlier saturation and a moderate reduction of the required undulator length. 

Such features are particularly attractive in the THz regime, where the FEL parameter is typically small, slippage effects are pronounced, and maintaining an efficient beam-radiation interaction is challenging. In addition, radiation diffraction becomes significant at long wavelengths, making it difficult to control the radiation size over extended undulator lengths and further emphasizing the benefits of compact, high-gain configurations.

Seeded-OK FELs~\cite{Kim1986OK, Yu1991, Yu2000,Huang2007} have been proposed to improved temporal coherence by imprinting a controlled modulation from an external laser. However, in the THz range this approach is limited by the lack of high-power, high-repetition-rate seed lasers synchronized to the electron beam. To overcome this limitation, alternative approaches based on laser frequency beating, chirped laser pulses, or two-color mixing have been proposed to generate long-wavelength intensity modulations suitable for THz FEL seeding \cite{Xiang2012, Malik2010,Zhang2024}.
In addition, oscillator optical klystron, where the optical klystron can also operate inside an FEL oscillator, avoids the need for an external laser, but the rapidly expanding optical mode in the THz regime severely reduces overlap with the electron beam during transport, limiting achievable gain ~\cite{Vinokurov1971,Coisson1981, Colson1985, Minehara1992, Takano1991}. 

This paper demonstrates that the unseeded optical klystron (OK) strategy leverages the natural THz slippage to synchronize the energy-modulation phase across the electron bunch. The dispersive chicane then converts this modulation into strong microbunching, and the radiator provides rapid coherent amplification without requiring any external seed, thereby enabling FEL operation at high repetition rates.
We investigate the performance of an unseeded optical klystron tailored to the THz regime using numerical simulations at resonant wavelengths of 10, 30, and 100~$\mu$m. 
We demonstrate how slippage affects energy modulation, bunching, and pulse growth, and we show the strategies such as harmonic bunching to mitigate large $R_{56}$ requirements. 
Finally, we propose and simulate a novel chicane-embedded optical delay scheme to compensate slippage and enable staged amplification, reaching multi-megawatt peak power. 
The results establish a pathway toward compact, high-intensity THz FEL sources based on the optical klystron principle.

\section{Optical Klystron Principle}
\label{sec:theory}
\subsection{concept}
The optical klystron is conceptually based on the klystron principle in microwave electronics \cite{Orzechowski1986,Huang2007}, 
in which an electron beam is first velocity-modulated in a cavity and then density-modulated in a drift space, producing coherent amplification in a second cavity.
In the optical analogue, the two cavities are replaced by undulator segments, and the drift space is replaced by a dispersive magnetic chicane.

In the first undulator, the interaction of the electron beam with the radiation field induces a small energy modulation at the resonant wavelength $\lambda_r$. For a beam with initial energy $E_0$ and modulation amplitude $\Delta E$, the relative energy deviation is \cite{Bonifacio1984FEL, Saldin2000}
\begin{equation}
    \delta = \frac{\Delta E}{E_0} \cos(k_r z),
\end{equation}
where $k_r = 2\pi/\lambda_r$ is the resonant wavenumber and $z$ is the longitudinal coordinate within the bunch.

The dispersive section, characterized by its momentum compaction $R_{56}$, converts this energy modulation into a longitudinal displacement
$\Delta z = R_{56} \, \delta .$
As a result, the longitudinal distribution of the beam acquires a periodic density modulation at the resonant wavelength. The strength of this modulation is quantified by the bunching factor $b = \left\langle e^{i k_r z} \right\rangle ,$
where the average is taken over the electron ensemble. 
In practice, the finite uncorrelated energy spread $\sigma_\delta$ of the electron beam reduces the achievable bunching, since the phase spread induced by $R_{56}\sigma_\delta$ smears out the modulation. The effective bunching factor including energy spread can be approximated as \cite{Penco2017,Yu1991, Saldin2000}
\begin{equation}
b \simeq J_1 \!\left(k_r R_{56} \frac{\Delta E}{E_0}\right) 
\exp\!\left[-\tfrac{1}{2}(k_r R_{56}\sigma_\delta)^2\right].
\label{eq:bunching}
\end{equation}
This expression illustrates the trade-off: while larger $R_{56}$ increases bunching, it also amplifies the detrimental effect of energy spread.

After the chicane, the uncorrelated energy
spread acts as a damping term and can be neglected if it is much
smaller than the energy modulation induced in the modulator
therefore the bunching factor can be expressed as \cite{Penco2017}
\begin{equation}
b \simeq J_1 \!\left(k_r R_{56} \frac{\Delta E}{E_0}\right),
\end{equation}
where $J_1$ is the first-order Bessel function of the first kind. This shows that the bunching oscillates as a function of the product $k_r R_{56} \Delta E / E_0$.
The condition for maximum bunching is obtained when the argument of the Bessel function corresponds to its first maximum.
Thus, the optimal dispersive strength is \cite{Penco2017}
\begin{equation}
R_{56}^{\mathrm{opt}} \approx \frac{1.84}{k_r} \frac{E_0}{\Delta E}.
\end{equation}
This expression highlights the interplay between the modulation amplitude in the first undulator and the required strength of the dispersive section. A stronger energy modulation allows a weaker $R_{56}$, while a weaker modulation requires a stronger dispersive section.

Therefore, in designing an optical klystron, $R_{56}$ must be optimized considering both the induced modulation and the intrinsic beam quality.

\begin{figure*}
    \centering
    \includegraphics[trim={0 4.5cm 0 3.cm},clip, width=16 cm,height=5.5 cm]{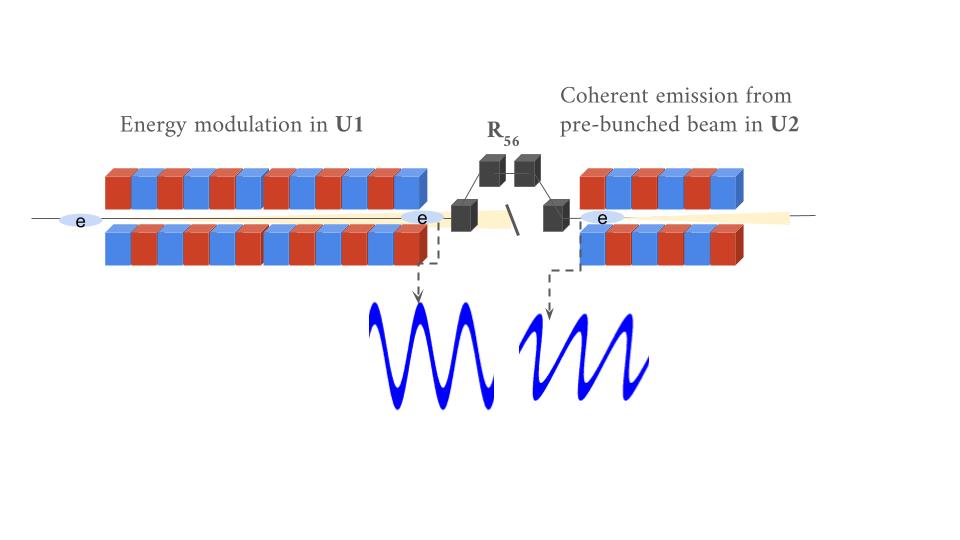}
    \caption{Optical klystron schematic with radiation dump between undulators.
In U1 the electron beam acquires energy modulation at \(\lambda_r\).
Radiation from U1 continues straight and is absorbed in a dump placed inside the chicane, while the electron beam is deflected by the dipoles.
The chicane (\(R_{56}\)) converts energy modulation into density modulation (microbunching), and the pre-bunched beam radiates coherently from the entrance of U2.}
\label{fig:OK_schematic_dump}

\end{figure*}

\subsection{Unseeded (SASE-type) Optical Klystron}

In the unseeded configuration, the optical klystron starts from shot noise: no external field is injected into the first undulator (U1), and the initial radiation arises from spontaneous emission. A key feature we exploit in the THz regime is the large \emph{slippage} between the radiation wave and the electrons in a sufficiently long modulator.
In an undulator, the radiation slips ahead of the electrons by approximately one resonant wavelength $\lambda_r$ per undulator period. Therefore, for a modulator with $N_u$ periods, the total slippage is
\begin{equation}
S \;\simeq\; N_u\,\lambda_r .
\end{equation}
When the electron bunch length $L_b$ is \emph{comparable to} the total slippage $S$,
\begin{equation}
L_b \;\sim\; S \;\; \Longleftrightarrow \;\; N_u \;\sim\; \frac{L_b}{\lambda_r},
\label{eq:slippage_condition}
\end{equation}
the optical wave emitted in the tail of the bunch samples a large fraction of the bunch, effectively \emph{phase-mixing} the interaction along $z$ within the bunch. In this regime, the magnetic undulator field imprints a longitudinally correlated energy modulation across the bunch, and the accumulated slippage tends to \emph{align the modulation phase} of head and tail sections despite the noise start. In other words, a \emph{long modulator} can partially synchronize the phase of the induced energy modulation across a THz-scale bunch (this phenomenal is illustrated in figure \ref{fig:bunching_phase} ).

Immediately downstream of U1, a magnetic chicane with momentum compaction $R_{56}$ converts this energy modulation into density modulation (microbunching). Because the microbunching is already partially phase-aligned by slippage in U1, the bunching factor at the entrance of the radiator (U2) is enhanced compared with a standard short SASE modulator. Radiation produced in U1 is removed \emph{on-axis inside the chicane} (mirror/absorber to dump the field), so that only the pre-bunched electrons seed U2. The radiator then amplifies coherent emission from the very start of the interaction.

For design and optimization, three practical conditions guide the unseeded OK:
\begin{enumerate}
\item \textbf{Slippage matching:} choose the modulator length such that $S \mathlarger> L_b$ [Eq.~\eqref{eq:slippage_condition}] to maximize phase synchronization across the bunch.
\item \textbf{Dispersive tuning:} select $R_{56}$ around the optimum
\[
R_{56}^{\mathrm{opt}} \approx \frac{1.84}{k_r}\frac{E_0}{\Delta E},
\quad k_r = \frac{2\pi}{\lambda_r},
\]
accounting for energy-spread suppression $\exp[-(k_r R_{56}\sigma_\delta)^2/2]$ to preserve bunching.
\item \textbf{THz-specific constraints:} in the THz regime, space-charge forces and the long cooperation length can degrade bunching in overly long modulators. In our formula, $\sigma_\delta$ represents the effective uncorrelated energy spread entering the bunching suppression term, including the initial slice energy spread and any additional contribution accumulated along the modulator (e.g. due to longitudinal space-charge effects).
Thus, $N_u$ of modulator (U1) should be large enough to realize $S \mathlarger{>} L_b$ but not so large that energy spread growth or space-charge debunching offsets the slippage benefit. More quantitatively, the uncorrelated energy spread reduces the bunching after the chicane by the factor
$exp\![-\left(k_r R_{56}\sigma_\delta\right)^2/2].$
Requiring this suppression to remain moderate, for instance
\begin{equation}
\exp\!\left[-\frac{1}{2}\left(k_r R_{56}\sigma_\delta\right)^2\right] \gtrsim 0.5,
\end{equation}
yields the condition
$k_r R_{56}\sigma_\delta \lesssim \sqrt{2\ln 2}.$
For a dispersive strength close to the optimal value ($R_{56}^{\mathrm{opt}} $),

this implies the practical constraint
\begin{equation}
\sigma_\delta \lesssim 0.64\,\frac{\Delta E}{E_0}.
\end{equation}
Therefore, we must keep the total uncorrelated energy spread $\sigma_\delta(N_u)$ below the bound above. This provides a quantitative guideline for selecting the modulator length in the THz unseeded optical klystron regime. A detailed analysis of the additional contributions accumulated along the modulator (e.g., due to collective effects) is beyond the scope of this work and will be presented in a future publication. 

\end{enumerate}
\subsection{Energy Modulation in the First Undulator (SASE Start-Up)}

In the optical klystron configuration without an external seed, the radiation field in the first undulator (U1) starts from shot noise and grows according to the self-amplified spontaneous emission (SASE) mechanism. The energy modulation imprinted on the electron beam is therefore governed by the exponential growth of the FEL eigenmode. The initial rms bunching factor per radiation wavelength is given by
\begin{equation}
    b_0 \simeq \frac{1}{\sqrt{N_\lambda}}, 
    \qquad 
    N_\lambda = \frac{I\,\lambda_r}{e c},
\end{equation}
where $I$ is the beam peak current and $\lambda_r$ the resonant wavelength.  
In the high-gain linear regime, the bunching factor grows as
\begin{equation}
    b(z) \simeq b_0 \, e^{z/(2L_g)},
\end{equation}
with $L_g$ the FEL power gain length. For the fundamental eigenmode, the scaled energy modulation $p = \Delta\gamma/(\rho\gamma)$ and bunching $b$ satisfy the relation
\begin{equation}
    |p| = \sqrt{3}\,|b|,
\end{equation}
so that the rms energy modulation after a modulator of length $L_1$ is
\begin{equation}
    \Delta\gamma_{\mathrm{rms}}(L_1) 
    \;\approx\; \rho \gamma \; \frac{\sqrt{3}}{\sqrt{N_\lambda}} 
    \exp\!\left(\frac{L_1}{2L_g}\right).
    \label{eq:deltagamma_sase}
\end{equation}

In 1D theory the gain length is
\begin{equation}
    L_g^{(1D)} = \frac{\lambda_u}{4\pi\sqrt{3}\,\rho},
\end{equation}
where $\lambda_u$ is the undulator period and $\rho$ the Pierce parameter.  
Note equation~\eqref{eq:deltagamma_sase} shows that the energy modulation scales with the normalized beam energy $\rho\gamma$, but is suppressed at start-up by the shot noise factor $1/\sqrt{N_\lambda}$. A finite undulator length $L_1$ allows exponential growth, but the modulation remains much smaller than $\rho\gamma$ unless $L_1 \gtrsim 2$--$3L_g$.  

In the THz regime, diffraction and slippage are particularly strong, reducing the effective growth rate and further limiting the energy modulation. Therefore, U1 must be long enough to ensure $\Delta\gamma_{\mathrm{rms}}$ reaches a useful fraction of $\rho\gamma$ for efficient density modulation in the following dispersive section.
\cite{Bonifacio1984,Kim1986,Xie1995,Saldin2000}
\section{Simulation Setup} \label{sec:semi}
The simulations are benchmarked to the beam and lattice conditions foreseen for the DALI (Dresden Advanced Light Infrastructure) THz FEL \cite{Mirian2025, DALI-CDR, Helm2023}. DALI is a compact accelerator-based light source currently under development at HZDR, targeting the generation of short, intense, and spectrally flexible THz radiation. The facility is based on a superconducting electron linac delivering electron beams in the energy range of several tens of MeV with high bunch charge, optimized for strong coherent radiation emission and optical klystron schemes in the THz regime.

In our simulation transverse beam optics are matched across U1-chicane-U2 to preserve emittance and overlap.
Table~\ref{tab:dali_params} lists the electron beam parameters and baseline parameters used in the simulations. 
\begin{table}[t]
\centering
\caption{Baseline DALI THz-OK simulation parameters.}
\label{tab:dali_params}
\begin{tabular}{l c}
\hline
\textbf{Beam energy} \(E_0\) & 50\, MeV \\
\textbf{Bunch charge} \(Q\) & 1 \,nC \\
\textbf{RMS bunch length} $\sigma_{z}$ & 120 $\mu$m \\
\textbf{Peak current} \(I_p\) & 1\,kA\, \\
\textbf{Norm. emittance} \(\varepsilon_{n,x/y}\) & \,10 mm\,mrad\, \\
\textbf{RMS rel. energy spread} \(\sigma_\delta\) & 0.002 \\
\textbf{Repetition rate} \(f_{\mathrm{rep}}\) & 1 MHz \\
\textbf{Target wavelength} \(\lambda_r\) & 10 - 100 \(\mu\)m\, \\
\textbf{U1\&U2 Undulator type} & planar \\
\textbf{U1\&U2 Undulator period} \(\lambda_u\) & 100 \,mm\, \\
\textbf{U1\&U2 Undulator parameter} \(K\) & \,0.97--4.4\, \\
\textbf{Magnetic gap} \(g\) & \,40-100 mm \\
\textbf{U1 length} \(L_{u1}\) & 4 m \\
\textbf{U2 length} \(L_{u2}\) & 2 m \\
\textbf{Chicane compaction} \(R_{56}\) &   \,0--1 cm\, \\
\textbf{Beta functions} \(\beta_{x/y}\) & 8/0.3 m \\
\textbf{Aperture (vacuum)} & 35 mm \\
\hline
\end{tabular}
\end{table}
The performance of the optical klystron is determined by the choice of undulator parameters, in particular the period length $\lambda_u$, the magnetic gap, and the achievable peak magnetic field $B_\mathrm{max}$ \cite{XFEL_TDR}. 
The radiation wavelength is determined by the FEL resonance condition, $\lambda_r = \frac{\lambda_u}{2 \gamma^2}\left(1 + \frac{K^2}{2}\right),$
where $\gamma$ is the relativistic Lorentz factor of the electron beam. The parameter $K$ is the dimensionless undulator strength parameter, related to the peak field by 
$ K = e B_\mathrm{max} \lambda_u /2\pi m_e c,$ 
where $e$ is the elementary charge, $m_e$ the electron mass, and $c$ the speed of light.
Thus, by varying the undulator gap $g$, one tunes the peak field $B_\mathrm{max}$ and the parameter $K$, which in turn adjusts the resonant wavelength $\lambda_r$.

For the present optical klystron simulations at DALI, an undulator period of $\lambda_u = 10$~cm was chosen. 
This relatively long period is well suited for operation in the THz regime, where radiation wavelengths of 10~$\mu$m to 100~$\mu$m are targeted.

\begin{figure}
    \centering
    \includegraphics{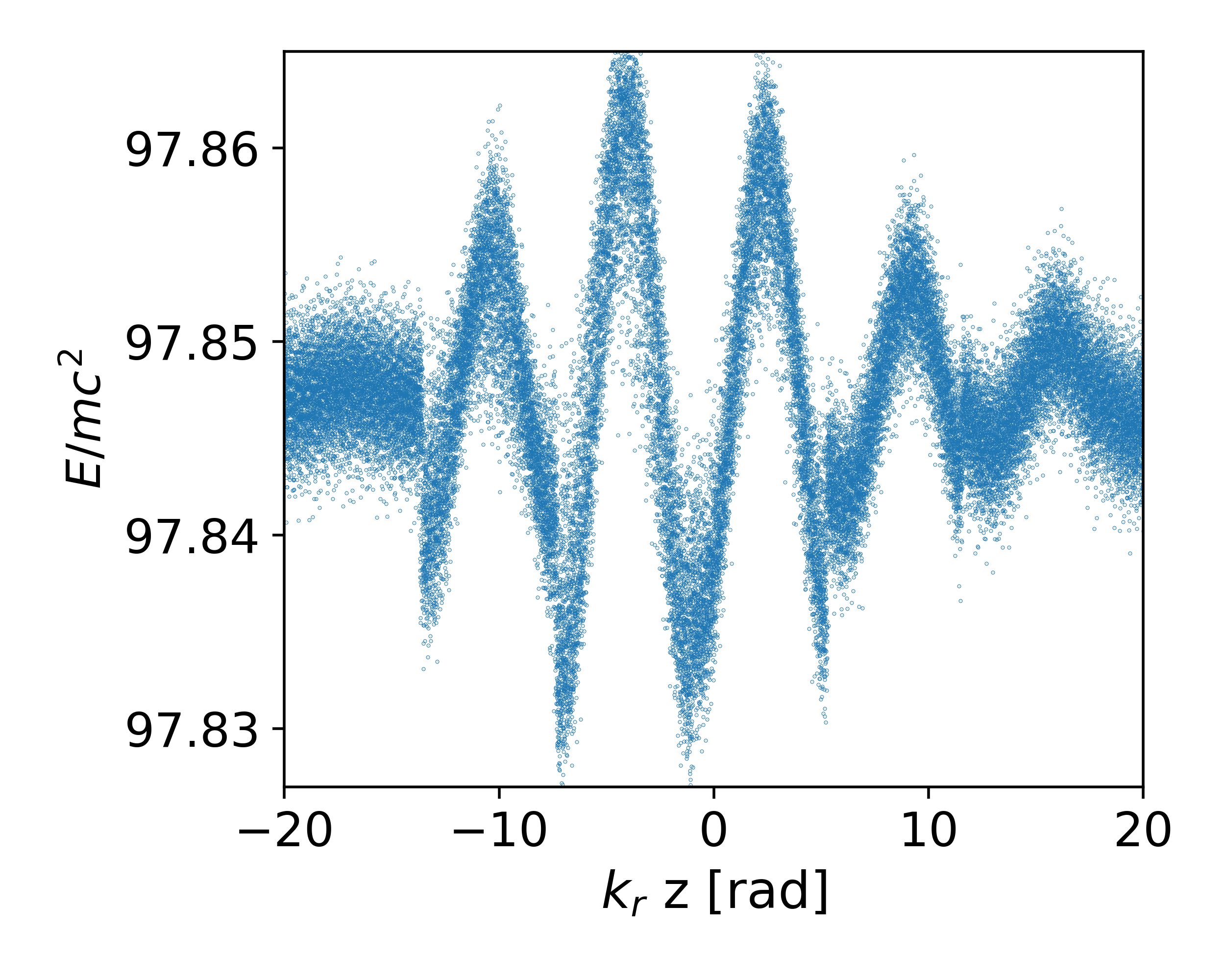}
    \caption{Simulated longitudinal electron phase space (energy vs.  $k_r~z$) obtained from time-dependent GENESIS simulations of the first undulator (U1) for a 1 kA electron bunch with 120 μm RMS length. The plotted longitudinal window spans approximately eight radiation wavelengths ($\lambda_r = 100~\mu$m), thereby covering the full bunch profile. Each slice is treated independently and macroparticles do not propagate between slices. As a result, the energy-phase correlation is continuous within each slice but not enforced to be continuous across slice boundaries. The apparent discontinuities therefore reflect the discrete slice representation and phase referencing, rather than a physical discontinuity in the beam energy modulation.} 
    \label{fig:particles}
\end{figure}
Simulations were carried out using \textsc{Genesis}~\cite{genesis}, including both shot-noise startup (SASE-like) and slippage effects. Three wavelengths were considered: 100~$\mu$m, 30~$\mu$m, and 10~$\mu$m. The modulator (U1) consists of 40 periods, and the electron bunch current profile was modeled as a Gaussian distribution with $\sigma = 120\,\mu\text{m}$. Figure~\ref{fig:particles} shows the resulting energy modulation at the resonance wavelength of 100~$\mu$m after U1. 
Here, the FWHM of the electron bunch is nearly three times the resonance wavelength, and the total slippage is 4 mm, it means more than 10 times of the FWHM of the electron bunch. 
The segmented structure observed in the longitudinal phase space arises from the multi-slice formulation of time-dependent GENESIS. Since macroparticles are confined to individual slices and no global longitudinal phase continuity is imposed, energy-phase correlations appear discontinuous when multiple slices are combined in a single plot.

For the present beam (\(E_0=\SI{50}{MeV}\), \(\sigma_\delta=0.002\)) and our simulated
first-undulator modulation shows relative modulation is $A = \Delta E/E_0 = 3.1\times 10^{-4}$ or in other words, \(A_\sigma= \Delta E/\sigma_\delta = 0.153\).
In the weak-modulation limit (\(A_\sigma<1\)), according to Eq.(\eqref{eq:bunching}) maximizing the fundamental bunching after
a dispersion section yields the well-known working point
\begin{equation}
  k R_{56}\,\sigma_\delta \simeq 1,
  \qquad\Rightarrow\qquad
  R_{56}^{\star} = \frac{1}{k\,\sigma_\delta} = \frac{\lambda_r}{2\pi\,\sigma_\delta}.
  \label{eq:R56_opt_weak}
\end{equation}

\begin{figure}
    \centering
    \includegraphics[width=5 cm]{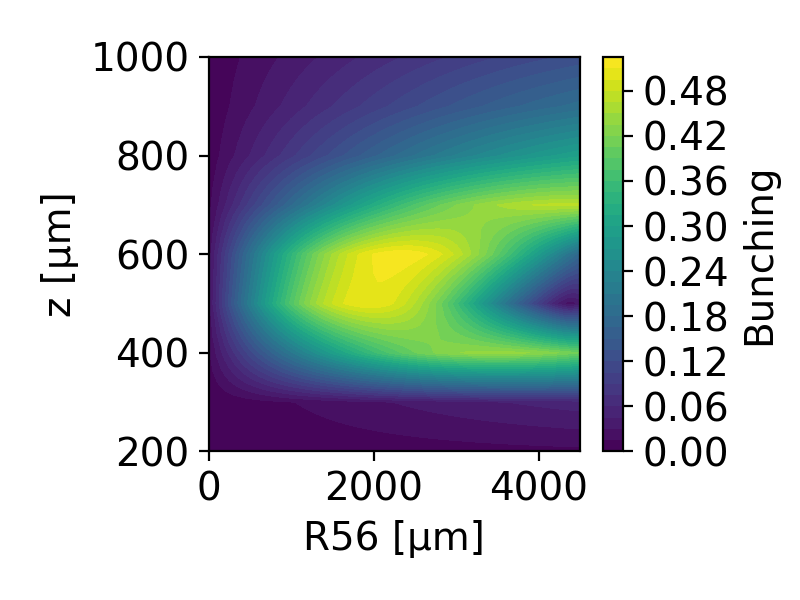 }
     \includegraphics[width=5 cm]{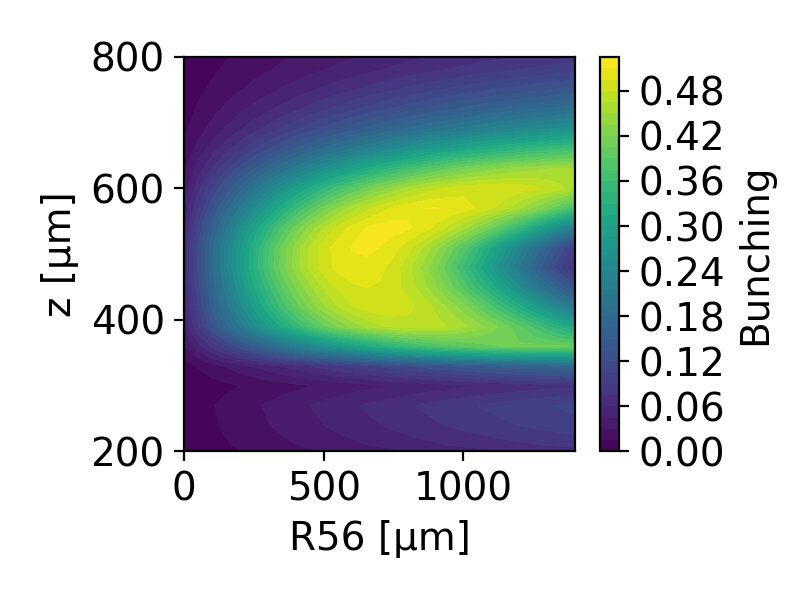 }
      \includegraphics[width=5 cm]{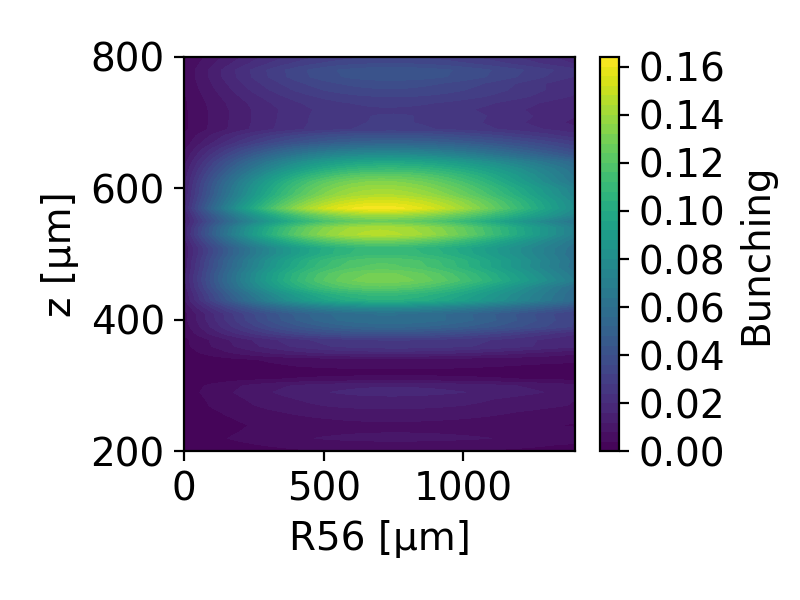 }
\caption{Bunching-factor maps $|b(z)|$ after the first chicane as functions of $R_{56}$ (horizontal axis) and longitudinal
position within the bunch $z$ (vertical axis) for $\lambda_r = 100~\mu$m (left), $30~\mu$m (middle), and $10~\mu$m (right).
For each $R_{56}$, $b(z)=\langle e^{i k_r z'}\rangle_{\text{slice}}$ is computed in longitudinal slices; the color scale indicates $|b|$.
The ridge of maxima marks the optimal dispersion; 
its shift to larger $R_{56}$ at $10~\mu$m
arises from the weaker energy modulation in U1.}

    \label{fig:Bunching_R56scan}
\end{figure}

We found that the induced energy modulation is nearly the same for a radiation wavelength of $30~\mu\text{m}$. For the $10~\mu\text{m}$ case, the simulation was carried out with an undulator parameter of $K = 0.97$. Due to the lower field strength, the resulting energy modulation was comparatively small, on the order of $2.6~\text{keV}$. Such a weak modulation requires larger values of $R_{56}$ to achieve significant bunching. 
Figure~\ref{fig:Bunching_R56scan} displays 2D maps of the bunching factor magnitude $|b(z)|$ after the first chicane,
plotted versus the dispersive strength $R_{56}$ (horizontal axis) and the longitudinal position within the bunch $z$
(vertical axis), for $\lambda_r = 100~\mu$m (left), $30~\mu$m (middle), and $10~\mu$m (right).
For each $R_{56}$ setting, the local complex bunching is evaluated in longitudinal slices according to
$b(z)=\langle e^{i k_r z'}\rangle_{\text{slice}}$, and the color scale encodes $|b|$. 
The ridge of maximal $|b|$ identifies the optimal dispersion: it shifts to larger $R_{56}$ at shorter wavelength
(e.g., $\sim 0.9$~mm at $10~\mu$m versus $\sim 0.4$~mm at $30~\mu$m). 
The vertical extent of the high-$|b|$ region reflects phase synchronization by slippage; it covers a larger fraction of the
bunch at $100~\mu$m, while at $10~\mu$m the modulation is more localized.

\begin{figure}
    \centering
    \includegraphics[width=6 cm]{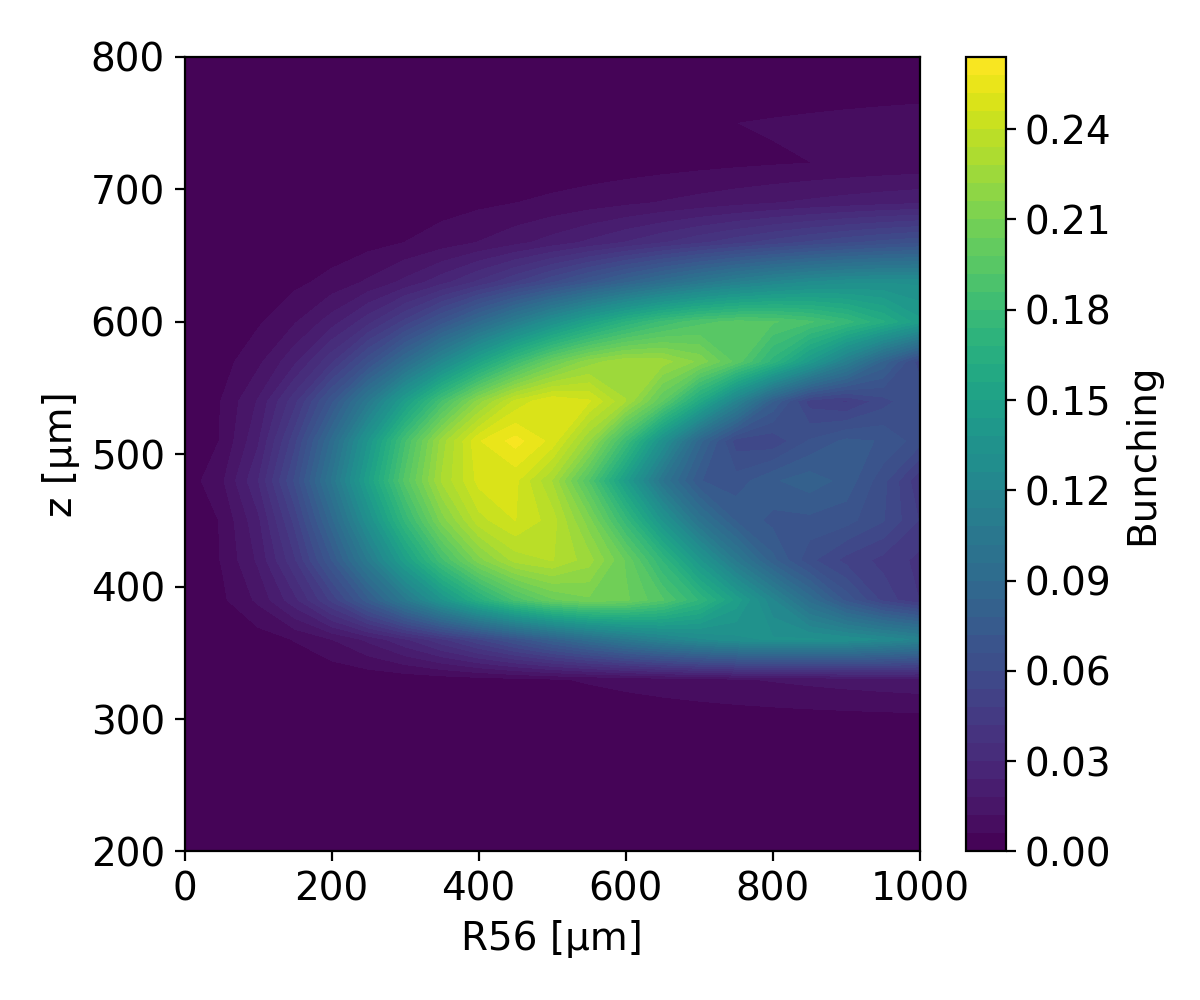}
    \caption{
Numerically evaluated maps of the third-harmonic bunching factor, $|b_{3}(z)|$, resulting from a fundamental energy modulation at $\lambda = 30,\mu\mathrm{m}$ after the first chicane. The bunching factor is shown as a function of the longitudinal dispersion $R_{56}$ (horizontal axis) and the longitudinal position within the electron bunch $z$ (vertical axis). The corresponding fundamental bunching factor at $30,\mu\mathrm{m}$ is shown in Fig.~3 (middle).
}
    \label{fig:3rdHarmonic}
\end{figure}

The results indicate that, for a radiation wavelength of $10~\mu\text{m}$, an $R_{56}$ value of approximately $900~\mu\text{m}$ is required to achieve significant bunching. An alternative strategy to avoid such a large dispersive strength is to exploit the third harmonic of the $30~\mu\text{m}$ energy modulation. As shown in Fig.~\ref{fig:3rdHarmonic}, with an $R_{56}$ of only $400~\mu\text{m}$, the bunching factor at $10~\mu\text{m}$ already reaches about $0.25$.

\begin{figure}
    \centering
    \includegraphics[width=6 cm]{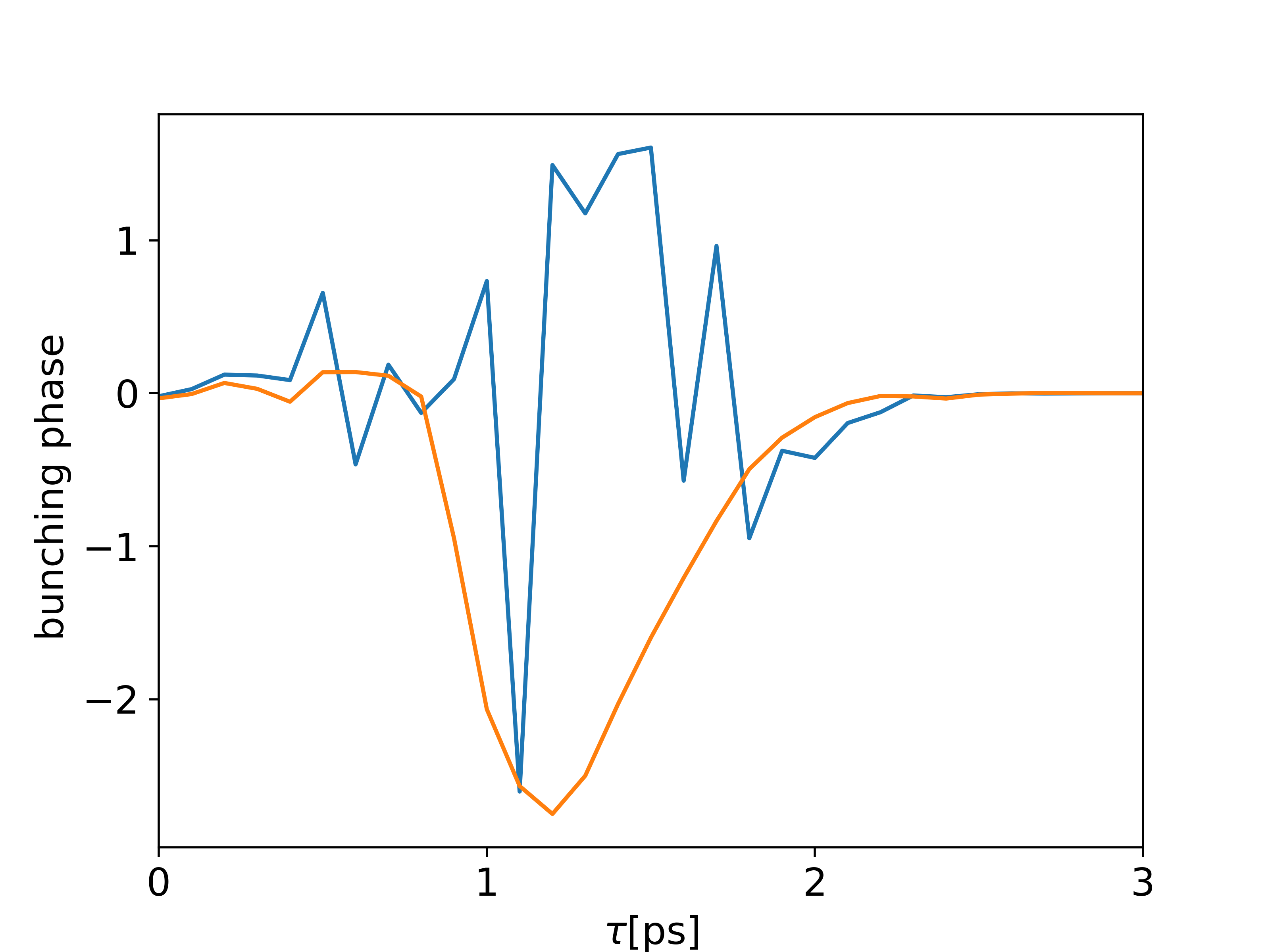}
    \caption{Phase of the complex bunching factor at 30 $\mu m$ as a function of the longitudinal coordinate after the first undulator period (blue) and after 40 undulator periods (orange).
At the early stage of the interaction, the bunching phase exhibits strong slice-to-slice fluctuations characteristic of shot-noise-dominated emission. After propagation over many undulator periods, radiation slippage couples neighboring slices, leading to a smoother and more correlated longitudinal phase profile across the electron bunch.  }
    \label{fig:bunching_phase}
\end{figure}

In order to illustrate the role of slippage in the THz regime, the phase of the complex bunching factor at a wavelength of 30 $\mu m$ is shown in figure \ref{fig:bunching_phase} as a function of the longitudinal coordinate after the first undulator period (blue) and after 40 undulator periods (orange), without applying any dispersive section. At the early stage of the interaction, the bunching phase exhibits strong slice-to-slice fluctuations characteristic of shot-noise-dominated emission. After propagation over many undulator periods, radiation slippage allows the electromagnetic field emitted by upstream slices to interact with downstream electrons, effectively coupling neighboring slices. As a result, the bunching phase evolves toward a smoother and more longitudinally correlated profile across the electron bunch.

\begin{figure}
    \centering
    \includegraphics[width=8 cm, height=5 cm]{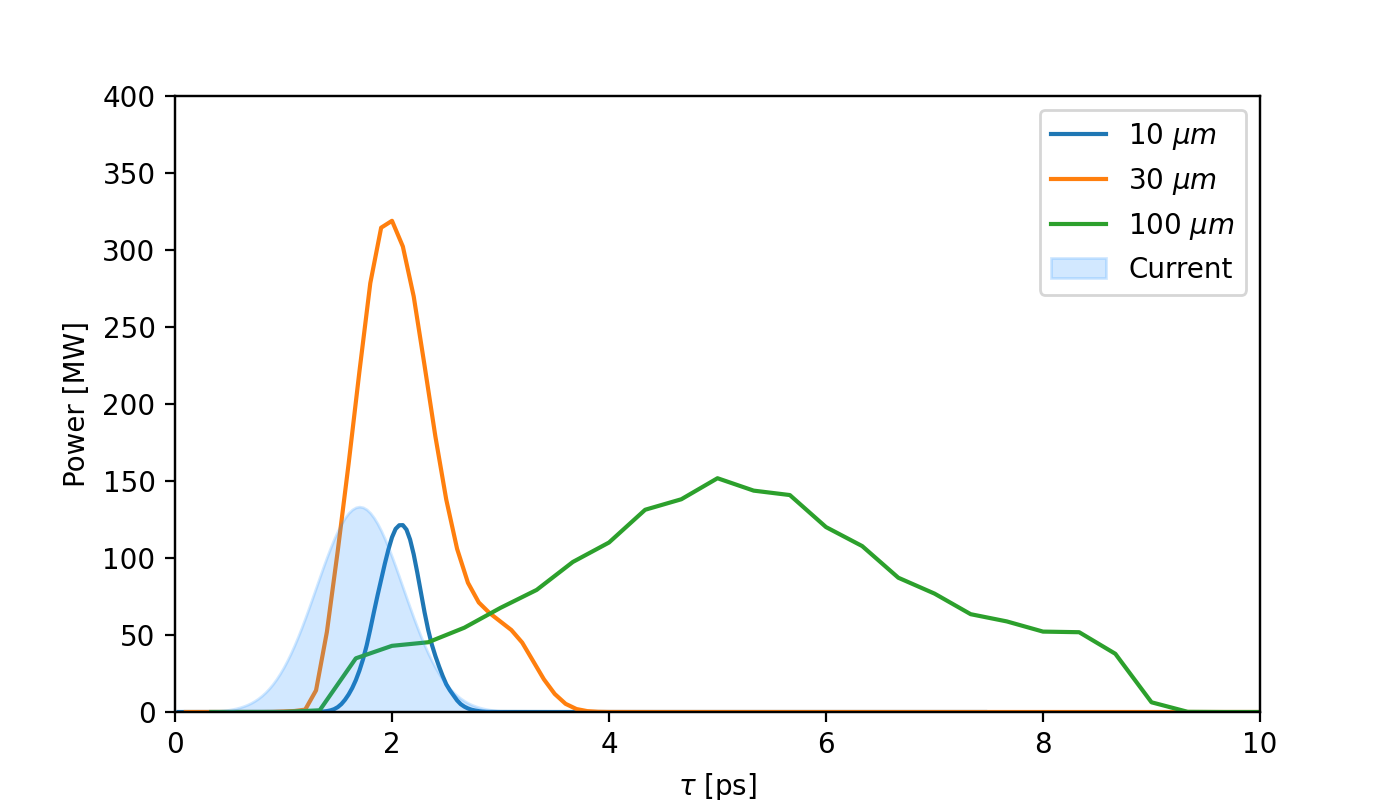} \includegraphics[width=8 cm, height=5 cm]{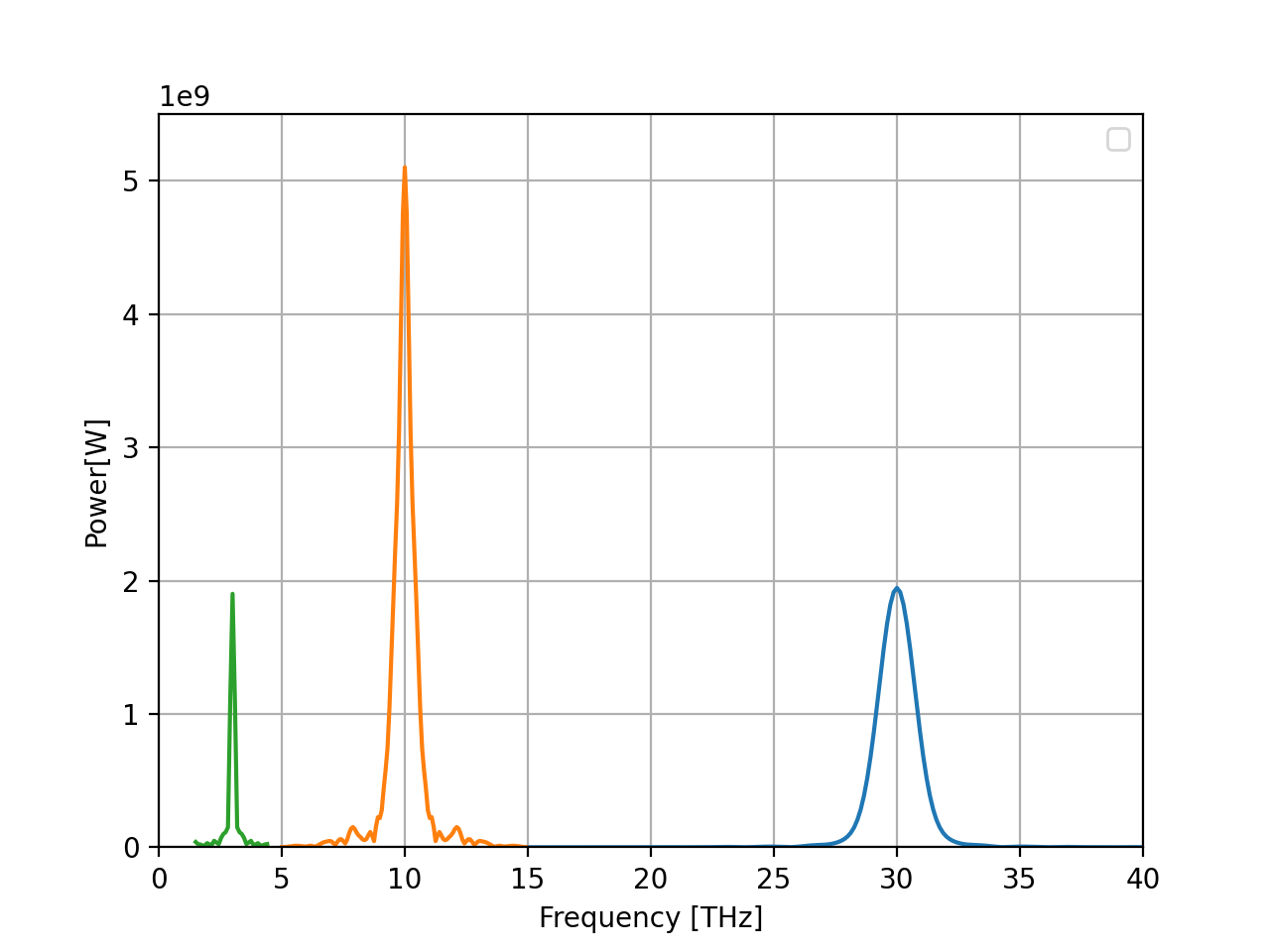}
    \caption{Left: Simulated FEL output power after the second undulator (U2) for radiation wavelengths of $10~\mu\text{m}$, $30~\mu\text{m}$, and $100~\mu\text{m}$. 
The blue shaded area represents the electron beam current profile. 
The effect of slippage is evident in the temporal structure of the FEL pulses, with longer wavelengths exhibiting broader pulse durations due to stronger slippage. Right:  Corresponding spectra, showing narrowband emission centered at the resonant frequencies for each wavelength.}

    \label{fig:FELpuls}
\end{figure}

After optimizing $R_{56}$ to achieve the desired bunching factor, the electron beam is directed into the second undulator (U2), which has the same parameters as U1 but with 20 periods instead of 40. Figure \ref{fig:FELpuls} demonstrates the simulated FEL pulse after U2.  
The blue shaded area represents the electron beam current profile, while the colored curves show the simulated FEL output power for radiation wavelengths of $10~\mu\text{m}$, $30~\mu\text{m}$, and $100~\mu\text{m}$. 
It should be emphasized that the $10~\mu\text{m}$ results presented above correspond to direct modulation of the electron beam in U1 at $10~\mu\text{m}$, rather than being generated as the third harmonic of the $30~\mu\text{m}$ modulation. The harmonic bunching analysis shown in Fig.~\ref{fig:3rdHarmonic} is included only as an alternative pathway to reduce the required $R_{56}$ for efficient bunching at short wavelengths.
The temporal overlap between the current profile and the FEL pulses illustrates the role of slippage: at shorter wavelengths, the radiation remains more confined within the high-current region of the bunch, whereas at longer wavelengths the increased slippage leads to a broader temporal distribution of the FEL pulse. This behavior highlights the challenge of maintaining strong interaction in the THz regime and underlines the importance of tailoring the optical klystron parameters to balance bunching strength and slippage effects. This figure demonstrates that via this approach enables the generation of coherent ultrashort THz pulses with durations of a sub picoseconds (FWHM) and peak powers in the multi-hundred-megawatt range at wavelengths 10 and 30 $\mu$m.
The corresponding spectra in the right panel of Fig.~\ref{fig:FELpuls} confirm narrowband emission centered at the resonant frequencies for all three wavelengths. The spectral width decreases with increasing wavelength, consistent with stronger slippage and longer coherence length at $100~\mu\text{m}$.

\begin{figure}
    \centering
    \includegraphics[width=8 cm]{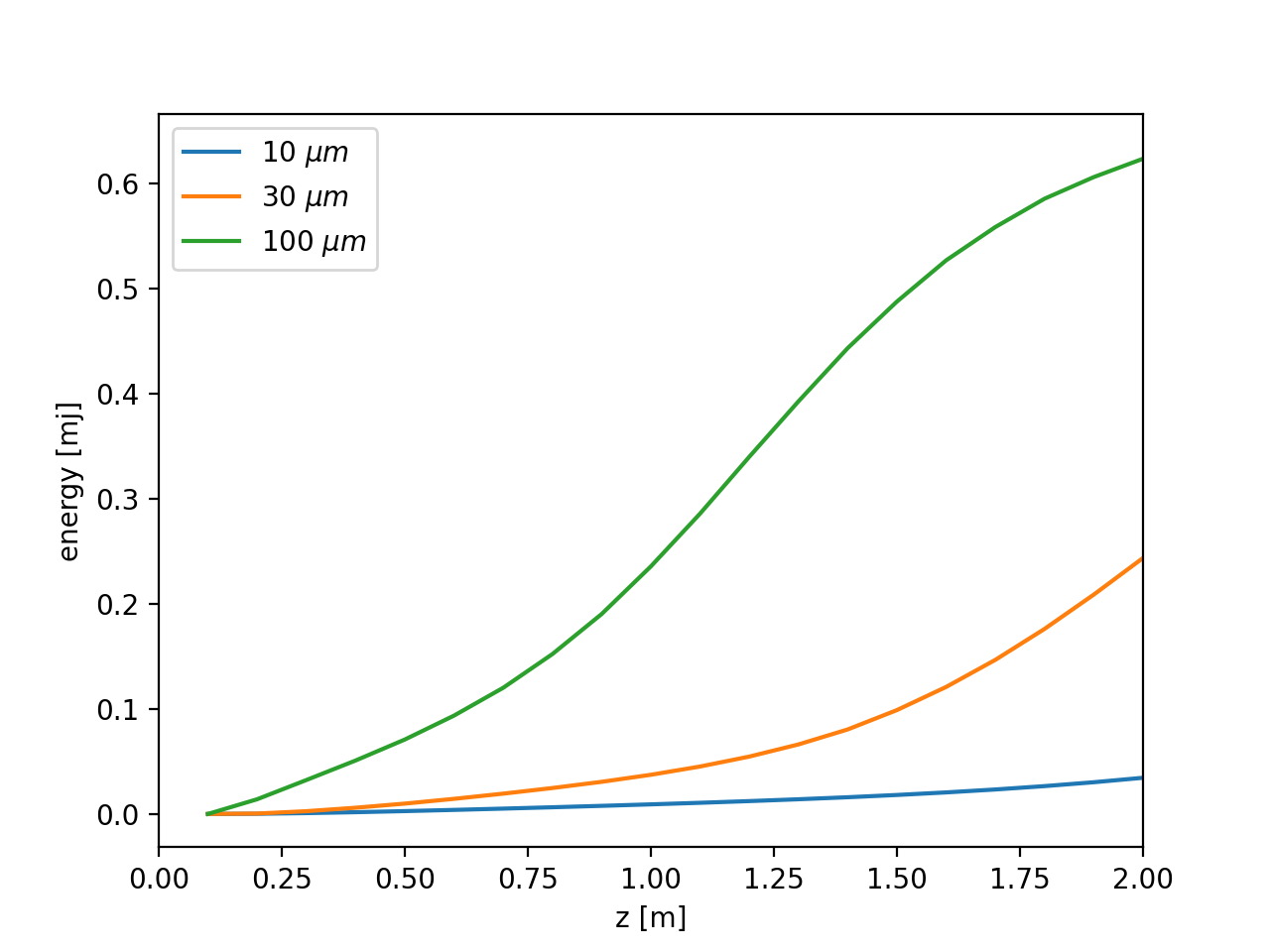}
    \caption{Simulated FEL pulse energy growth along the undulator for radiation wavelengths of $10~\mu\text{m}$, $30~\mu\text{m}$, and $100~\mu\text{m}$. The $100~\mu\text{m}$ case shows the fastest growth, exceeding $0.6$~mJ at $2$~m, whereas the shorter wavelength cases exhibit slower energy buildup.}

    \label{fig:PulseEnergy}
\end{figure}

As a complementary view, Fig.~\ref{fig:PulseEnergy} shows the evolution of the FEL pulse energy along the undulator length for radiation wavelengths of $10~\mu\text{m}$, $30~\mu\text{m}$, and $100~\mu\text{m}$. 
The longer wavelength case ($100~\mu\text{m}$) exhibits the fastest energy growth, reaching more than $0.6$~mJ after $2$~m of undulator length, while the shorter wavelength cases show correspondingly slower growth ($~50 \,\mu$J) due to weaker coupling and stronger sensitivity to energy spread and slippage effects.

\begin{figure}
    \centering
    \includegraphics[width=8 cm]{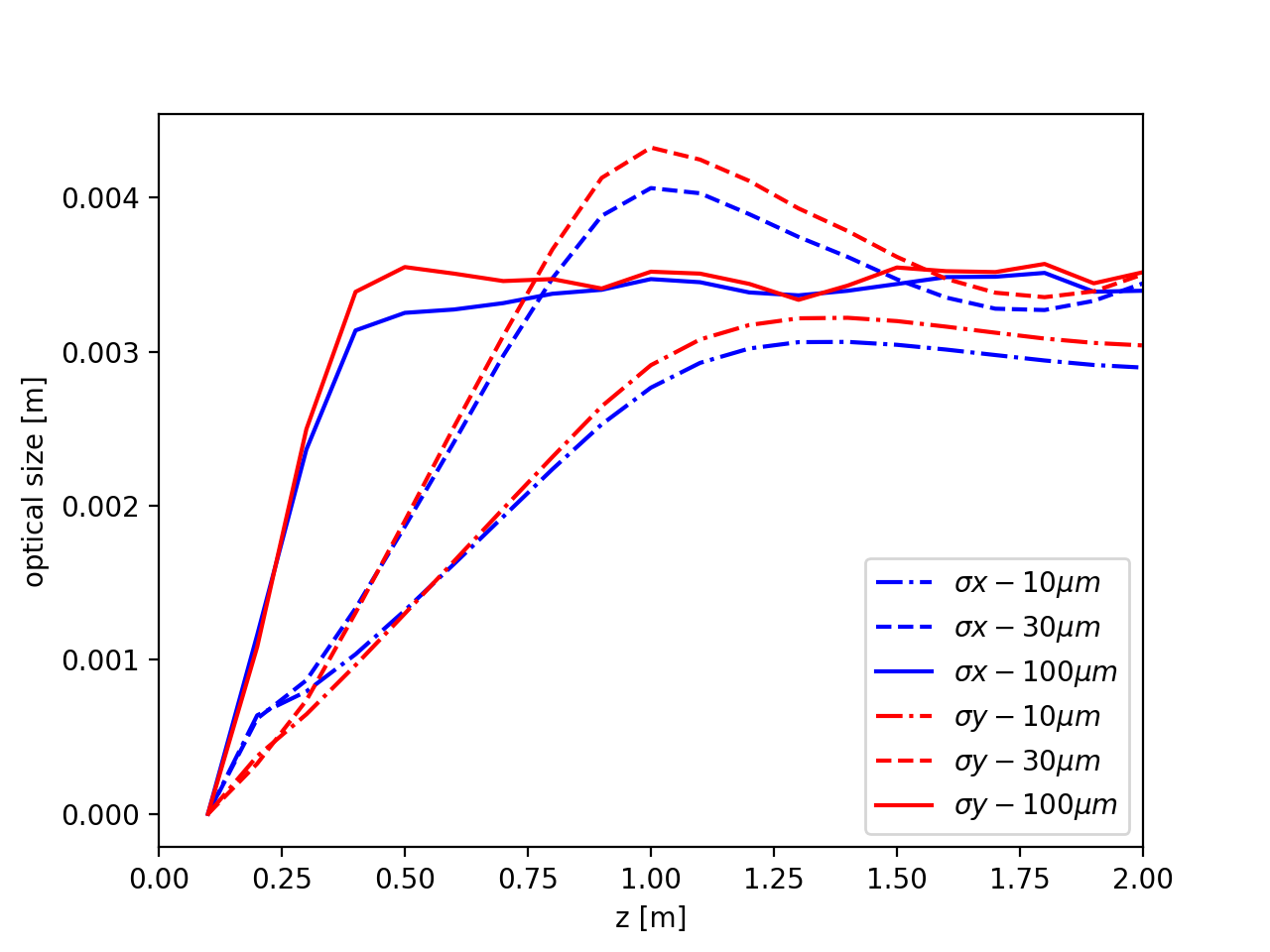}
   \caption{Evolution of the optical mode size ($\sigma_x$, $\sigma_y$) along the undulator for radiation wavelengths of $10~\mu$m, $30~\mu$m, and $100~\mu$m. 
Although the radiation spot size increases rapidly due to diffraction, it remains safely within the $3.5$~cm beamline aperture.}

    \label{fig:OpticalSize}
\end{figure}
As discussed earlier, diffraction of THz radiation presents a serious challenge because of the short Rayleigh length at long wavelengths. 
To quantify this effect, we tracked the transverse radiation size in both the horizontal ($\sigma_x$) and vertical ($\sigma_y$) planes along the undulator, as shown in Fig.~\ref{fig:OpticalSize}. 
For all three simulated wavelengths ($10~\mu$m, $30~\mu$m, and $100~\mu$m), the optical mode expands rapidly within the first meter and then saturates at values around $3$--$4$~mm. 
Importantly, the radiation spot size remains well below the available beamline aperture of $3.5$~cm, ensuring that diffraction does not lead to significant radiation losses in the present design. 
Nevertheless, the observed growth illustrates the inherent difficulty of maintaining high overlap between the electron beam and the radiation field, particularly at the longest wavelength.

In summary, the simulations confirm that an optical klystron can significantly enhance microbunching and FEL output in the THz regime. 
The performance depends strongly on wavelength: while longer wavelengths benefit from faster energy growth, shorter wavelengths require harmonic bunching or large dispersion. 
Diffraction, although present, remains below aperture limits and is therefore not the primary constraint in the present design. 

\begin{figure}
    \centering
    \includegraphics[trim={0 4cm 0 3cm},clip, width=16 cm,height=5.5 cm]{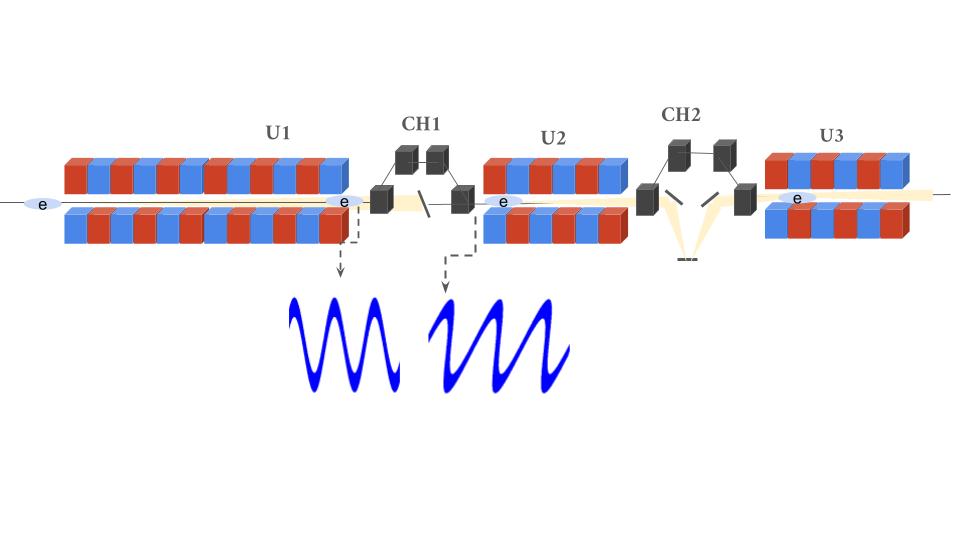}
    \caption{Schematic of the proposed optical klystron with chicane-embedded optical delays. 
The first undulator (U1) generates energy modulation, which is converted into density modulation by the chicane dispersion CH1. 
An optical delay line inside the second chicane CH2 retards the radiation wavefront, compensating the slippage accumulated in U2 and restoring overlap with the pre-bunched electrons in U3. 
A second stage ( chicane + U3) further amplifies the radiation with repeated re-phasing.}

    \label{fig:OKdelay}
\end{figure}

The electron beam parameters used in this study (50 MeV beam energy, 1 nC bunch charge, and rms bunch length of 120 $\mu$m) correspond to the design target values of the DALI facility and represent an ambitious but well-motivated operating regime for intense THz generation and Optical klystron concepts. Achieving such parameters at low beam energies is technically demanding due to collective effects such as space charge and coherent synchrotron radiation. However, similar regimes have been explored experimentally at several accelerator facilities, demonstrating the feasibility of high-charge (hundreds of pC to nC-class) and sub-picosecond electron bunches in the tens-of-MeV energy range, although not necessarily simultaneously in a single operating mode \cite{Zhang2017}. The present study aims to assess the performance and robustness of the proposed optical klystron scheme under these challenging, yet realistic, design conditions.

\section{Alternative Strategies for Mitigating Slippage Effects}\label{sec:alternatives}
As illustrated by our simulations, one of the main challenges in extending the optical klystron concept to the THz regime is the strong radiation slippage relative to the electron bunch. 
For long wavelengths such as $100~\mu\text{m}$, the interaction benefits from rapid energy growth, but the accumulated slippage leads to significant temporal broadening of the FEL pulse. 
Addressing this slippage challenge is therefore essential for realizing a compact and efficient THz optical klystron.

Several alternative strategies can be considered to mitigate the impact of slippage. 
One promising strategy to handle the slippage challenge is to introduce an \textbf{optical delay line embedded in the chicanes} between successive undulator sections. 

The strategy is as following, illustrated in Fig.~\ref{fig:OKdelay}, the first undulator (U1) imprints an energy modulation on the electron beam, which is converted into density modulation and enhanced in the first chicane (CH1) through its longitudinal dispersion $R_{56}$. 
The microbunched beam then emits coherently in the second undulator (U2). 
Due to the accumulated slippage in U2, the radiation field advances with respect to the electrons by
\begin{equation}
\Delta s_{\mathrm{slip}} \simeq N_u \lambda_r ,
\end{equation}
where $N_u$ is the number of undulator periods and $\lambda_r$ is the resonant wavelength. 

The second chicane (CH2) introduces additional longitudinal dispersion, which further enhances the density modulation according to
$\Delta z = R_{56}\,\Delta\gamma /\ \gamma$,
and, at the same time, increases the electron path length, thereby producing an additional delay of the electron beam with respect to the radiation field. 
An adjustable optical delay line accommodated within CH2 delays the radiation field by a controllable longitudinal shift $\Delta s_{\mathrm{od}}$, which is chosen to compensate both the slippage accumulated in U2 and the additional electron delay introduced by CH2. 
As a result, the electrons and radiation enter the third undulator (U3) in optimal phase, enabling efficient interaction and strong amplification from the beginning of U3.

This configuration effectively decouples the two roles of the chicane: 
(i) providing longitudinal dispersion to generate density modulation, and 
(ii) serving as a platform for an optical delay line that restores temporal alignment between the radiation packet and the electron bunch. 
The combination enables efficient multi-stage growth of microbunching even at long wavelengths where slippage would otherwise wash out the interaction. 
While the detailed engineering of such compact THz optical delay lines requires dedicated study, the concept can be implemented using a simple multi-mirror delay-line geometry, for example based on three- or four-mirror arrangements commonly employed in ultrafast and FEL beamlines \cite{DelayLine, NewportXMStage}. 
Such configurations allow continuous adjustment of the optical path length while preserving the beam direction and transverse mode quality, and therefore offer a practical route to compensating slippage effects in optical klystrons.

\begin{figure}
    \centering
    \includegraphics[width=6 cm]{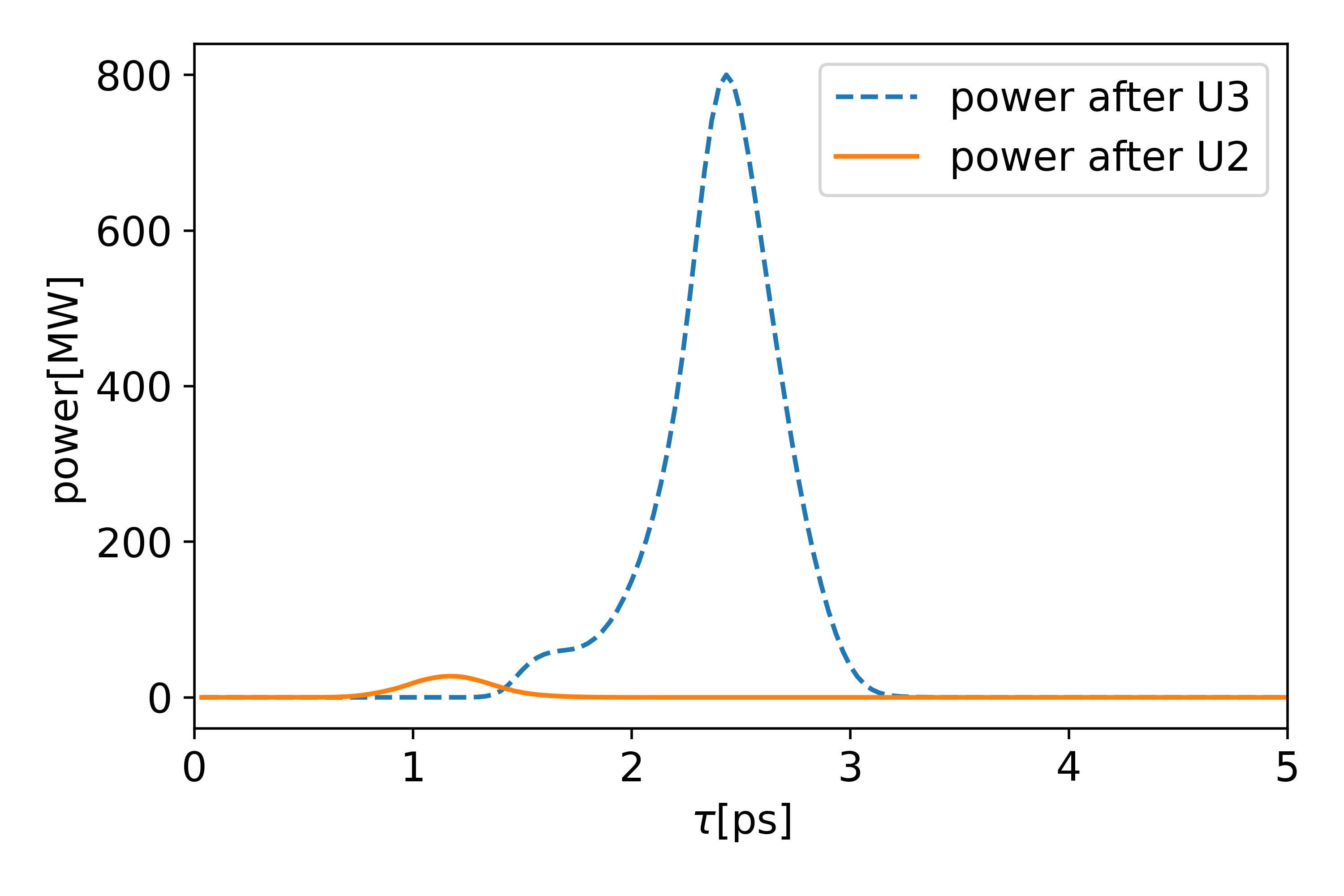}
    \includegraphics[width=6 cm]{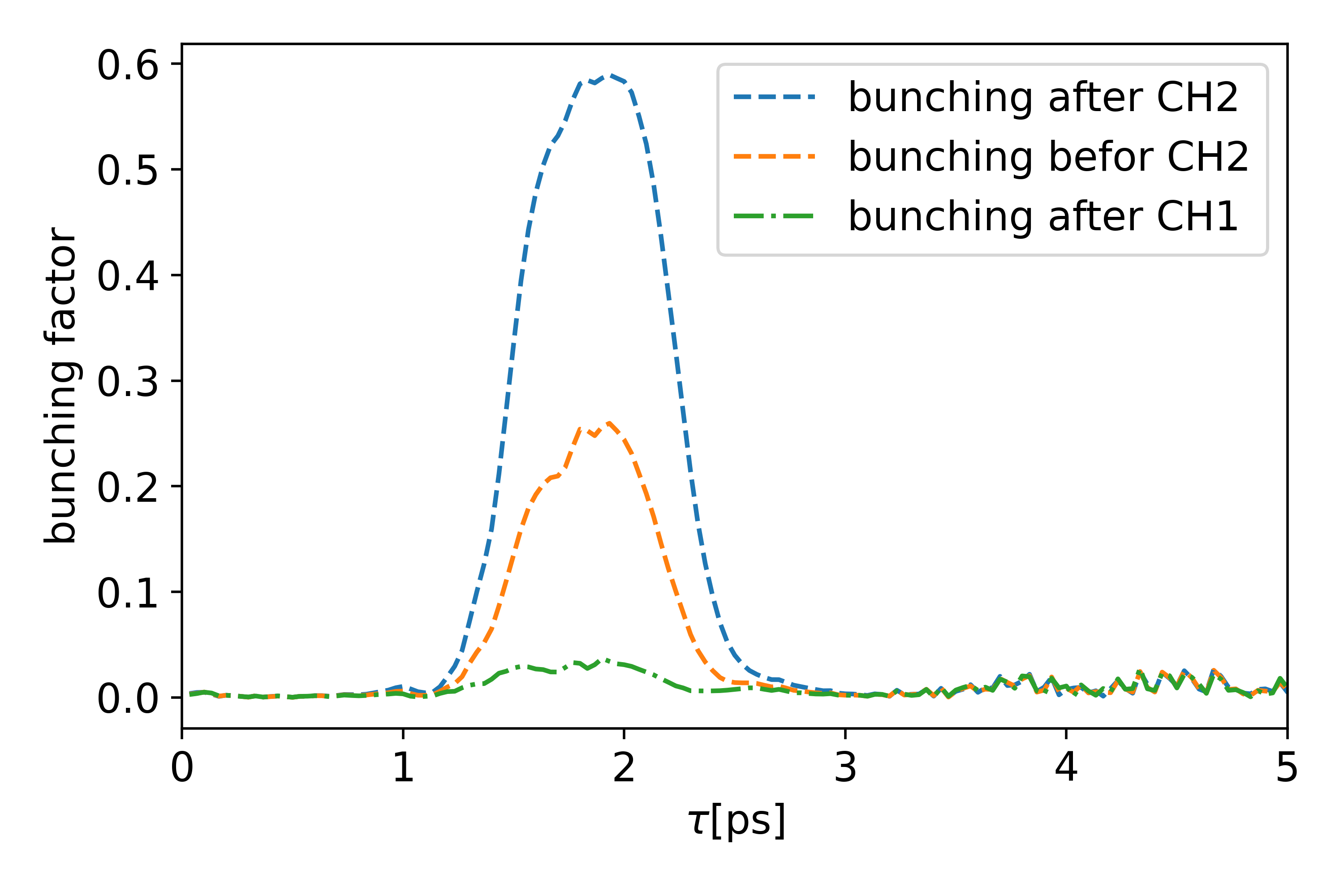}
    \caption{Simulated performance of the staged optical klystron with chicane-embedded optical delay. 
Left: FEL output power after U1 and U2. 
Right: bunching factor evolution at different positions: after first chicane (CH1), before second chicane (CH2), and after CH2. 
A moderate $R_{56}$ in the first chicane limits energy extraction in U2 but preserves beam quality, enabling strong amplification in U3.}

\label{fig:R56strategy}
\end{figure}

Numerical simulations confirm that careful adjustment of the dispersive strengths in the two chicanes is essential for efficient operation of the proposed multi-stage optical klystron. 
If a large $R_{56}$ is applied in the first chicane (CH1), strong microbunching is generated and U2 produces high radiation power. 
However, this rapid energy extraction effectively exhausts the beam quality, leaving little headroom for amplification in the subsequent undulator U3. 
A more favorable strategy is to employ a smaller $R_{56}$ in CH1, which generates only weak coherent radiation in U2 but preserves beam quality for the second stage, as illustrated in Fig.~\ref{fig:R56strategy}.

In figure~\ref{fig:R56strategy}, we present simulations of $10~\mu\mathrm{m}$ wavelength radiation using an optical delay-line configuration. 
The simulation parameters of electron beam and first and second undulators are summarized in Table~\ref{tab:dali_params}. 
The parameters of the third undulator (U3) are identical to those of the second undulator (U2), except that the length of U3 is $3~\mathrm{m}$. 
To optimize the radiation power, longitudinal dispersions of $R_{56}=2.0~\mathrm{mm}$ for CH1 and $R_{56}=120~\mu\mathrm{m}$ for CH2 are used. 
With these parameters, the radiation power at the exit of U2 remains below $100~\mathrm{MW}$. 
In the second stage, comprising the optical delay line, chicane CH2, and undulator U3, the delayed radiation interacts efficiently with the pre-bunched electron beam in U3, resulting in a peak power of up to $800~\mathrm{MW}$ with a pulse duration of approximately 500 fs (FWHM).
To clarify the underlying mechanism, the evolution of the bunching factor is shown at three locations: after the first chicane (CH1), before the second chicane (CH2), and after CH2, illustrating the progressive enhancement of microbunching through dispersion and re-phasing.

Another promising approach is \textbf{waveguide-assisted propagation}. 
The use of parallel-plate or corrugated waveguides can significantly extend the Rayleigh length of THz radiation, thereby reducing diffraction and helping to maintain overlap between the electron beam and the optical field. 
A waveguided undulator section or transport line suppresses the effective slippage by confining the radiation mode and enhancing its on-axis field strength. 
Such waveguide concepts have already been investigated in the context of THz FEL oscillators~\cite{Carr2002,Yang2024,Miginsky2015} and could be adapted to optical klystron configurations to improve efficiency and compactness.
The implementation of waveguide-assisted optical klystrons requires a dedicated study of mode propagation, beam-wave coupling, and cavity or transport integration. 
A comprehensive investigation of these aspects lies beyond the scope of the present work and will be addressed in future studies.

\section{Conclusion and Outlook}
\label{sec:conclusion}
We have investigated the feasibility of extending the optical klystron (OK) concept into the THz regime, where strong radiation slippage and diffraction present fundamental challenges. Using numerical simulations, we analyzed the performance of an OK at wavelengths of 10 $\mu$m, 30 $\mu$m, and 100 $\mu$m. The results show that while long wavelengths exhibit rapid energy growth, they suffer from significant temporal broadening due to slippage. Shorter wavelengths, in contrast, require large dispersive strengths $R_{56}$ or harmonic bunching strategies to achieve sufficient microbunching. Diffraction was also studied and found not to be a limiting factor in the present design, as the optical spot size remains well within the available aperture.

To address the slippage challenge, we discussed alternative strategies. We proposed and numerically demonstrated a novel chicane-embedded optical delay scheme, where controlled optical delays are introduced to re-phase the radiation with the electron bunch between undulator stages. Simulations confirm that careful tuning of the dispersive strengths in successive chicanes is crucial: using a moderate $R_{56}$ in the first chicane preserves beam quality and enables strong amplification in the final stage, reaching power levels of several megawatts.  Waveguide-assisted propagation can extend the Rayleigh length and maintain beam-wave overlap, though its implementation requires a dedicated study beyond the scope of this work.

In addition to slippage and diffraction, other collective effects-such as longitudinal space-charge forces, coherent synchrotron radiation in the chicanes, and undulator wakefields-are expected to play an important role in the THz regime. Their quantitative impact on bunching preservation and radiation growth must be included in future simulation studies to obtain a complete performance assessment. The dispersive section amplifies the influence of energy spread, which can smear out the density modulation and reduce the bunching factor \cite{mirian2021, mirian2025prab}.  In the THz regime, space-charge forces and the relatively long cooperation length further complicate this balance. As a result, while the optical klystron has strong potential for improving FEL performance, its successful implementation requires precise optimization of both beam and lattice parameters.

In summary, the optical klystron remains a promising approach for generating short, high-intensity THz pulses. By tailoring the design to explicitly account for slippage, dispersion, diffraction, and collective effects, compact and efficient THz FEL sources can be realized. The strategies outlined here provide a foundation for future experimental studies and for the development of advanced THz radiation facilities.

\section*{Acknowledgments}

The authors gratefully acknowledge fruitful discussions and support from colleagues at Helmholtz-Zentrum Dresden-Rossendorf (HZDR) and partner institutions. 
In particular, we thank Ulf Lehnert for valuable input and guidance throughout this work, Atoosa Meseck (HZB) and Michael Klopf (HZDR) for insightful comments on optical klystron strategies.
We also thank the accelerator physics groups at HZDR for their constructive feedback during the development of this study.

\nocite{*}

\bibliographystyle{unsrt}   
\bibliography{main}

\end{document}